\documentclass[reprint,
showpacs,preprintnumbers,
amsmath,amssymb,
aps,
pra]{revtex4-2}
\usepackage[dvipsnames]{xcolor}
\usepackage{graphicx}
\usepackage{braket}
\usepackage{amsmath}
\usepackage{hyperref}

\usepackage[T1]{fontenc}

\DeclareUnicodeCharacter{2212}{-}

\begin{document}


\title{Steady-State Emission of Quantum-Correlated Light in the Telecom Band from a Single Atom}%

\author{Alex Elliott}

\author{Scott Parkins}
\affiliation{%
Te Whai Ao--Dodd-Walls Centre for Photonic and Quantum Technologies}
\affiliation{%
Department of Physics, University of Auckland, Auckland 1010, New Zealand.}
\author{Takao Aoki}
\affiliation{%
Department of Applied Physics, Waseda University, 3-4-1 Okubo, Shinjuku-ku, Tokyo 169-8555, Japan
}
\date{\today}

\begin{abstract}
We propose and investigate a scheme for the \textit{steady-state} emission of quantum-correlated, telecom-band light from a single multilevel atom. By appropriately tuning the frequency of a pair of lasers, a two-photon transition is continually driven to an atomic excited state that emits photons at the desired wavelength. We show that resonantly coupling a cavity mode to the telecom transition can enhance the rate of emission while retaining the antibunched counting statistics that are characteristic of atomic light sources. We also explore coupling a second, independent cavity mode to the atom, which increases the telecom emission rate and introduces quantum correlations between the cavity modes. A model for the hyperfine structure of a single cesium atom is then described and numerically integrated to demonstrate the viability of implementing the scheme with a modern cavity QED system.

\end{abstract}

\maketitle


\section{Introduction}
Stable long-distance transmission of electromagnetic signals requires low-loss waveguides and associated technologies optimized to the carrier wavelength. Sources of quantum light in the telecommunications band are therefore particularly valuable, given the advanced infrastructure for their transmission and processing. To optimally leverage single-atom-based sources of quantum light, it is therefore critical to identify transitions that emit at a suitable wavelength. In general, this means identifying transitions between a pair of unstable excited states that are separated by an appropriate energy difference. 

Alkali atoms such as rubidium and cesium are a stalwart of modern quantum optics experiments due, in part, to their particularly simple energy level structures. Notably, both atoms have experimentally accessible transitions that are resonant in the telecom band. In atomic cesium, the $6P_J\leftrightarrow7S_{1/2}$ transitions have wavelengths of $\lambda = 1.36\mu\text{m}$ and $\lambda = 1.47\mu\text{m}$, for $J=1/2$ and $J=3/2$, respectively \cite{Toh2018, Toh2019}. In atomic rubidium, the $5P_J\leftrightarrow4D_{3/2}$ transition wavelengths are $\lambda = 1.48\mu\text{m}$ and $\lambda = 1.53\mu\text{m}$, for $J=1/2$ and $J=3/2$, respectively \cite{Moon2009, SteckRb87}. Although the focus in this paper is on a model of atomic cesium, the developed concepts are also pertinent to an implementation with rubidium.

The inherent delay between successive photon emissions from an atomic transition makes for promising single-photon sources in the telecommunications band \cite{Menon2020,Covey2023}. Additionally, consecutive (i.e., cascaded) decays from a higher-lying atomic state have shown promise as a source of correlated photons in a pair of cavity modes \cite{Chiarella2024}. However, achieving \emph{steady-state} emission of light on these transitions is impeded by parasitic spontaneous emission, given that neither state in the transition is stable. Overcoming this complication requires some form of optical pumping that can control the atomic population accordingly. Specifically, alkali atoms have two hyperfine ground levels, and care must be taken to avoid trapping the population in the wrong ground state. In this paper, we therefore propose and theoretically validate a scheme that yields a \emph{steady-state} emission of telecom-band quantum light, using a single atom. This is achieved with a pair of lasers that drive a two-photon ladder transition from one of the hyperfine ground levels, while simultaneously repumping population that leaks into the other, non-resonant ground state. The two-photon excitation proceeds via one level of the $P$ manifold fine-structure doublet, and we explore the emission of light that follows from relaxation through the other fine-structure component. Coupling a cavity mode to the privileged telecom transition is demonstrated to enhance and channel the emitted light, while preserving the quantum correlations that are characteristic of atomic emission. Furthermore, we show that coupling the subsequent optical (or near-IR) transition to a second, independent cavity mode can enhance the emission rate from the telecom cavity, and establish nonclassical two-mode correlations between the cavities.   

The cavities that we consider could be formed from two pairs of fiber Bragg-gratings, printed on either side of a section of tapered nanofiber \cite{Kato2015, Nayak2019, Horikawa2025}. Bragg gratings are tailored to highly-specific wavelengths, so one that is reflective in the telecom band will be essentially transparent in the optical band, and vice versa. Therefore, in principle, one can engineer two cavities with significantly disparate frequencies around a single piece of nanofiber. This would allow an atom held in the evanescent field of the nanofiber to simultaneously interact with independent optical and telecom cavity modes that have output fields directly into fiber, ideal for long distance transmission. While nanofiber cavities are particularly desirable for this context, the specific cavity architecture is not critical to the scheme, and other appropriate configurations, such as crossed fiber-tip Fabry-P\'erot cavities \cite{Brekenfeld2020, Chiarella2024} could work similarly well in practice.

The paper is organized as follows. In Section \ref{Conceptual Model}, we outline a conceptual five-level, ``diamond''-configuration model of an atom that qualitatively illustrates key operating principles and features of the scheme. Factors that affect the rate and statistics of photon emissions are explained by calculating these properties in both the one- and two-mode cavity output fields. Section~\ref{Coupling to the 7s State in Practice}, then describes implementation of the diamond model with a cesium atom, incorporating the full hyperfine structure of the relevant energy levels. We calculate the photon counting statistics of the cavity output fields and show that the system can behave as a steady-state source of quantum-correlated light, violating one-mode and two-mode Cauchy-Schwarz inequalities. 

\section{Conceptual Model}
\label{Conceptual Model}
\begin{figure}[t]
 	\centering
 	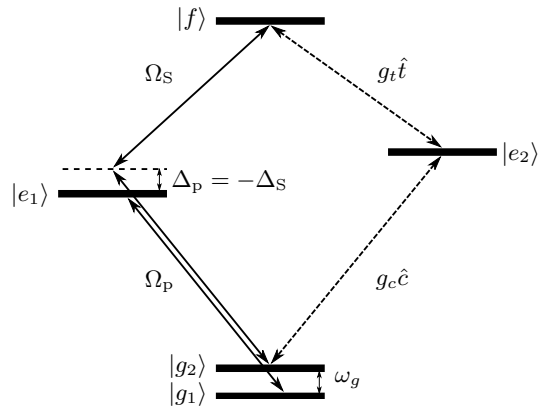
 	\caption{Schematic of the conceptual double-diamond model, with two driving lasers, configured such that the pump laser ($\Omega_{\rm p}$) is resonant with the $\ket{g_1}\leftrightarrow \ket{e_1}$ transition ($\Delta_{\rm p} = \omega_g$). The pump and Stokes ($\Omega_{\rm S}$) laser frequencies combine to be two-photon-resonant with the $\ket{g_2}\leftrightarrow \ket{f}$ transition, i.e. $\Delta_{\rm p} +\Delta_{\rm S} =0$. Solid arrows show classical coherent driving fields, and dashed arrows indicate coupling to quantized cavity modes. Ideally, the four fields implement a cycle of transitions from the $\ket{g_2}$ state, starting with a two-photon transition through $\ket{e_1}$, followed by a sequence of cavity-enhanced relaxations via $\ket{e_2}$.}
 	\label{fig1}
 \end{figure}

We begin by outlining a conceptual double-diamond model, as shown in Fig. \ref{fig1}, which is inspired by alkali atomic level structure and demonstrates the key operating principles of the scheme. The model atom has a total of five energy levels, two of which are the stable ground states $\ket{g_1}$ and $\ket{g_2}$. There are two intermediate states, $\ket{e_1}$ and $\ket{e_2}$, which independently couple to the pair of ground states in a V-type configuration. The $\ket{e_1}$ state couples and decays to both ground states in equal proportion, while $\ket{e_2}$ only couples and decays to the $\ket{g_2}$ state. Finally, a single excited state, $\ket{f}$, couples to the two intermediate states in a $\Lambda$-type configuration. We assume that the excited state spontaneously decays independently and symmetrically to the two intermediate states. Quantitative results relevant to a realistic implementation with a cesium atom are given in the following section, along with a discussion of suitable hyperfine levels that can be employed to mimic the double-diamond model. 

One branch of the double-diamond configuration is used for two-photon excitation from the ground states to the $\ket{f}$ state, via the $\ket{e_1}$ intermediate state. A pump laser ($\Omega_{\rm p}$) drives the $\ket{g_{1,2}}\leftrightarrow\ket{e_1}$ transitions and a Stokes laser ($\Omega_{\rm S}$) drives the $\ket{e_1}\leftrightarrow\ket{f}$ transition. The ground state splitting, $\omega_g$, is assumed to be large compared to both of the laser Rabi frequencies, and therefore a two-photon transition will only appreciably proceed from one of the two ground states. However, if the pump laser is near-resonant from either ground state, the population will eventually be dissipatively transferred to the other ground state. We therefore propose a scheme that is configured to utilize both the one- and two-photon resonances simultaneously. The pump laser is tuned near resonance with $\ket{g_1}\leftrightarrow\ket{e_1}$ , while the Stokes laser is configured, in combination with the pump, to give a two-photon resonance on the $\ket{g_2}\leftrightarrow\ket{f}$ transition. This arrangement of lasers gives control over the ground-state population distribution, and simultaneously provides a mechanism to excite $\ket{f}$ in steady state.

The second branch of the double-diamond configuration provides an alternative, independent pathway for the excited state to decay. Coupling suitable cavity modes to these transitions enhances their rate, which encourages the atom to preferentially decay through this branch. In practice, the  $\ket{f}\rightarrow\ket{e_2}$ transition is resonant in the telecommunications band, while the $\ket{e_2}\rightarrow\ket{g_2}$ transition is at a higher frequency. We refer to these as telecom and control cavities, respectively; the telecom cavity captures and channels atomic emission at the desired wavelength, while the control cavity aids in the overall cycle rate and establishes two-mode correlations that reflect the sequential transitions.

The atom-cavity system is described in an interaction picture, with the Hamiltonian in the rotating-wave approximation given by
\begin{equation}
\begin{split}
    \hat{H}/\hbar =&~-\omega_g\ket{g_1}\bra{g_1}+\Delta_\text{p}\ket{e_1}\bra{e_1} + (\Delta_\text{p}+\Delta_\text{S})\ket{f}\bra{f}\\
    &+\frac{\Omega_\text{p}}{2}\big(\hat{D}_\text{p}^\dagger + \hat{D}_\text{p}\big)+\frac{\Omega_\text{S}}{2}\big(\hat{D}_\text{S}^\dagger + \hat{D}_\text{S}\big)\\
    &+g_t\big(\hat{D}_t^\dagger\hat{t} +\hat{t}^\dagger \hat{D}_t\big)+g_c\big(\hat{D}_c^\dagger\hat{c} +\hat{c}^\dagger\hat{D}_c \big),
\end{split}
\end{equation}
where $\hat{t}$ and $\hat{c}$ are annihilation operators for the telecom and control cavity modes, respectively, and we have introduced electric-dipole transition operators,
\begin{equation}
    \begin{split}
        \hat{D}_\text{p} = \ket{g_1}\bra{e_1} +\ket{g_2}\bra{e_1},~\hat{D}_\text{S} = \ket{e_1}\bra{f}, \\
        \hat{D}_{t} = \ket{e_2}\bra{f},~\text{and}~\hat{D}_{c} = \ket{g_2}\bra{e_2}.\\
    \end{split}
\end{equation}
Time evolution of the composite density matrix, $\hat{\rho}$, is described by a Lindblad master equation of the form
\begin{equation}
\begin{split}
    \dot{\hat{\rho}} =\frac{1}{i\hbar}\left[\hat{H},\hat{\rho}\right]&+2\kappa_t\mathcal{D}\big(\hat{t}\hspace{0.1em}\big)+2\kappa_c\mathcal{D}\big(\hat{c}\big) \\
    +~ \gamma_\text{p}\mathcal{D}\big(\hat{D}_\text{p}&\big)+\gamma_\text{S}\mathcal{D}\big(\hat{D}_\text{s}\big)+\gamma_t\mathcal{D}\big(\hat{D}_t\big)+\gamma_c\mathcal{D}\big(\hat{D}_c\big),
    \end{split}
\end{equation}
where the dissipation superoperators are defined by $\mathcal{D}(\hat{o}) \equiv \hat{o}\hat{\rho}\hat{o}^\dagger -\frac{1}{2}\hat{o}^\dagger\hat{o}\hat{\rho}-\frac{1}{2}\hat{\rho}\hat{o}^\dagger\hat{o}$, for an operator $\hat{o}$. The telecom and control cavities have halfwidths $\kappa_t$ and $\kappa_c$, respectively. For simplicity, all unstable states have the same total linewidth $\gamma$, with the individual rates scaled to reflect the respective branching ratios, such that $2\gamma_\text{p} = 2\gamma_\text{S} =2\gamma_t =\gamma_c =\gamma$. The steady-state density matrix is found as a stationary solution of the master equation, satisfying $\dot{\hat{\rho}}_\text{ss}=0$. All numerical results within this manuscript are calculating using QuTiP \cite{qutip5}.
 
\subsection{Steady-State Atomic Populations}
\begin{figure}[h]
 	\centering
 	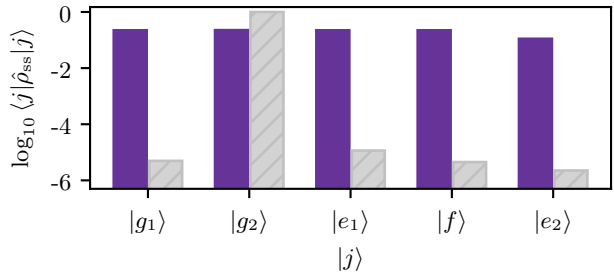
 	\caption{Log of steady-state atomic populations with a ground-state splitting $\omega_g = 1000\gamma$, and a symmetric laser driving ($\Omega_\text{p} = \Omega_\text{S}$). (Purple) The population is distributed across all the energy levels with the choice $\Delta_\text{p}=-\omega_g$ and $\Delta_\text{S}=\omega_g$, and with an effective two-photon transition rate $\Omega_\text{eff} \approx \Omega_\text{p}\Omega_\text{S}/\Delta_\text{S}=4\gamma$. (Grey hatched) Population becomes almost entirely confined in the $\ket{g_2}$ state, when $\Delta_\text{p}=\Delta_\text{S}=0$ with an equivalent two-photon transition rate $\Omega_\text{eff} \approx \sqrt{2\Omega_\text{p}\Omega_\text{S}}=4\gamma$.}
 	\label{fig2}
 \end{figure}

The appropriate one- and two-photon resonance conditions are achieved with a pump laser that is resonant with the $\ket{g_1}\leftrightarrow\ket{e_1}$ transition, and a Stokes laser tuned below the $\ket{e_1}\leftrightarrow\ket{f}$ transition frequency by the ground state splitting, $\omega_g$. Configuring the driving lasers accordingly, Fig.~\ref{fig2} shows the steady-state population for an atom in free space (no cavities). For comparison, the populations are also shown for lasers arranged to resonantly drive both one- and two-photon transitions from the $\ket{g_1}$ state, with an equivalent effective two-photon transition rate, demonstrating the population confinement in the doubly-off-resonant $\ket{g_2}$ state.

With just the two appropriately-tuned driving lasers in Fig.~\ref{fig2}, the atom continuously emits light on the telecom transition, at a rate $\gamma/2\braket{f|\hat{\rho}_\text{ss}|f}\approx 0.11\gamma$. The spontaneous emission is perfectly antibunched, as expected from an ideal two-state emitter, albeit with a flux that is restricted by the limited excited-state population and decay rate. The telecom emission rate cannot be further increased with stronger driving lasers, because the two-photon transition is essentially saturated, and the branching ratios of the $\ket{f}$ state are fixed. Instead, a solution that can enhance the telecom flux is to couple the transition to a near-resonant cavity mode.

\begin{figure}[tb]
 	\centering
 	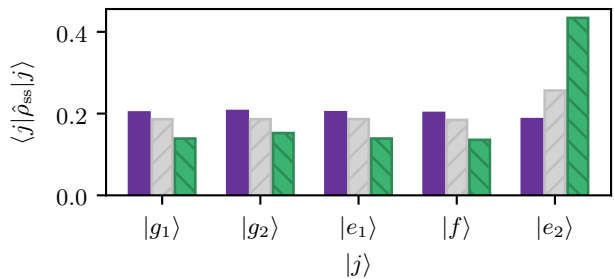
 	\caption{Steady-state populations for the atom coupled to a telecom cavity mode, with a ground-state splitting $\omega_g = 1000\gamma$, symmetrically driven ($\Omega_\text{p} = \Omega_\text{S}$) with an effective two-photon transition rate $\Omega_\text{eff} \approx \Omega_\text{p}\Omega_\text{S}/\Delta_\text{S}=4\gamma$, and $\Delta_\text{p}=-\Delta_\text{S}=-\omega_g$. Each case has a single-atom cooperativity $C_t = 8$, with $g_t = \{1,2,4\}\gamma$ ($\kappa_t=\{0.5,2,8\}\gamma$) for the purple, grey hatched, and green hatched bars, respectively.}
 	\label{fig3a}
 \end{figure}

For the purpose of increasing the flux, the utility of coupling a cavity mode to the telecom transition is two-fold; the total transition rate is enhanced and consequently the fraction of parasitic emission to the $\ket{e_1}$ state is suppressed. This is demonstrated in Fig.~\ref{fig3a}, which shows the steady-state atomic populations for the atom coupled to a telecom cavity mode on the $|f\rangle\leftrightarrow |e_2\rangle$ transition, for various cavity parameter sets, each set having the same single-atom cooperativity, defined as 
\begin{equation}
    C_t = \frac{2g_t^2}{\kappa_t\gamma_t},
\end{equation}
where $\gamma_t=\gamma /2$. As the telecom cavity coupling is increased, so too is the rate of transitions to the $\ket{e_2}$ state, which hence accumulates a larger fraction of the steady-state atomic population. A commensurately larger flux is channeled through the cavity into its output field as the coupling becomes stronger. The steady-state cavity output flux is calculated with the input-output theorem \cite{Collett1985}, assuming a vacuum cavity input field, as 
\begin{equation}
    \Phi_\text{ss}^t = 2\kappa_t\braket{\hat{t}^\dagger\hat{t}}_\text{ss},
\end{equation}
where $\braket{\hat{o}}_\text{ss}$ denotes the steady-state expectation value of operator $\hat{o}$. For example, with cavity parameters $g=2\kappa=\gamma$, one finds $\Phi^t_\text{ss}=0.086\gamma$, and for $2g=\kappa=8\gamma$, $\Phi^t_\text{ss}=0.366\gamma$.

Demonstrably, coupling a cavity mode to the relevant transition in the model atom is a means to channel telecom light in steady state. However, regardless of how strong the coupling is, the total output flux is still ultimately limited by free-space spontaneous emission. In particular, once a photon has been emitted from the telecom cavity, a subsequent emission is only possible after the $\ket{e_2}$ state has decayed. For strong coupling, reliance on this free-space decay ultimately becomes the rate-limiting step in the cycle of transitions, so population accumulates in the $\ket{e_2}$ state, as can be seen in Fig. \ref{fig3a}, for the stronger couplings. In principle, one can use a third laser to vacate the $\ket{e_2}$ state, but the associated Rabi oscillations will impede the ensuing two-photon transition step. Instead, we consider coupling the atom to a second, control cavity mode, resonant with the $\ket{e_2}\leftrightarrow\ket{g_2}$ transition. 

\begin{figure}[tb]
 	\centering
 	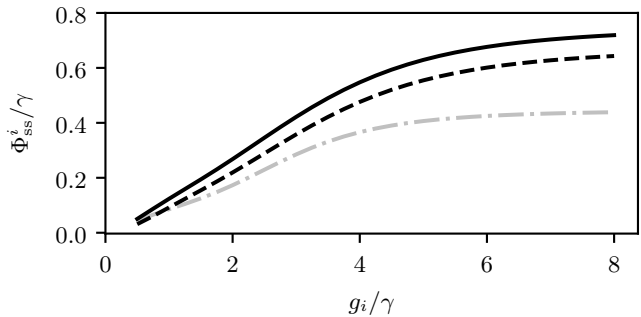
 	\caption{Steady-state output fluxes from the telecom (solid black) and control (dashed black) cavities, plotted against coupling strengths, $g_t=g_c=g_i$. The atom has a ground-state splitting $\omega_g = 1000\gamma$, and is symmetrically driven ($\Omega_\text{p} = \Omega_\text{S}$) with an effective two-photon transition rate $\Omega_\text{eff} \approx \Omega_\text{p}\Omega_\text{S}/\Delta_\text{S}=4\gamma$, and $\Delta_\text{p}=-\Delta_\text{S}-\omega_g$. In dot-dashed grey, the telecom cavity flux from an otherwise equivalent system in the absence of a control cavity is also shown. The decay rate is chosen so that the telecom single-atom cooperativity is always $C_t = 8$, and, where included, $\kappa_c=\kappa_t$.}
 	\label{fig4}
 \end{figure}

If the control cavity decay rate is suitably large, the photonic loss offers an essentially unidirectional coupling by avoiding Rabi cycling. Coupling to a suitably lossy control cavity is thus a favorable choice because the overall rate of the cycle of transitions can be enhanced without interrupting the two-photon excitation process. Generally, parameters for the control cavity should be similar to those for the telecom cavity, to give an optimal enhancement in the flux. To be concise, we therefore focus on the case where both cavities have identical parameters. With this choice, Fig. \ref{fig4} shows the dependence of telecom and control cavity fluxes on their respective coupling strengths, with a constant single-atom cooperativity. This shows the efficacy of using a control cavity to further enhance the telecom flux beyond the capacity of the single-cavity system. At large coupling strengths, the flux saturates at a value that is limited by the effective two-photon driving strength, rather than by any of the atomic linewidths. 

\subsection{Second-Order Correlations}
Single-atom light sources are well-studied emitters of light that has nonclassical photon correlations. Coupling a cavity to an atomic transition is a means to capture and channel the quantum-correlated light, under appropriate conditions. Here, we calculate the photon correlations from the telecom and control cavities, to predict their structure and time-dependence. Second-order correlation functions are used to quantify the likelihood of detecting a photon at a time interval $\tau$ before or after a trigger detection. There are two distinct classes of steady-state correlation function that we calculate, using quantum regression formulae \cite{Carmichael1999}. Firstly, the auto-correlation function relates to sequential detections from the same field mode,
\begin{equation}
    g^{(2)}_{\hat{o},\,\text{ss}}(\tau) = \frac{\braket{\hat{o}^{\dagger}(0)\hat{o}^{\dagger}(\tau)\hat{o}(\tau)\hat{o}(0)}_\text{ss}}{\braket{\hat{o}^{\dagger}\hat{o}}_\text{ss}^2},
\end{equation}
for a mode operator $\hat{o}$. Secondly, the cross-correlation function pertains to the conditional likelihood of detecting a photon in one mode, given a detection in the other,
\begin{equation}
    g^{(2)}_{\hat{o}_1\hat{o}_2,\,\text{ss}}(\tau) = \frac{\braket{\hat{o}_1^{\dagger}(0)\hat{o}_2^{\dagger}(\tau)\hat{o}_2(\tau)\hat{o}_1(0)}_\text{ss}}{\braket{\hat{o}_1^{\dagger}\hat{o}_1}_\text{ss}{\braket{\hat{o}_2^{\dagger}\hat{o}_2}_\text{ss}}},
\end{equation}
where $\hat{o}_1$ and $\hat{o}_2$ are annihilation operators corresponding to the modes of interest.

Coincidence is one aspect of second-order correlations that quantifies the likelihood that two photons are detected simultaneously. Cauchy-Schwarz inequalities impose limitations on the allowable values that can be achieved with a classical field for the auto- and cross-correlations:
\begin{equation}
    g^{(2)}_{\hat{o},\,\text{ss}}(0)\geq1,~
    g^{(2)}_{\hat{o}_1\hat{o}_2,\,\text{ss}}(0)\leq \sqrt{g^{(2)}_{\hat{o}_1,\,\text{ss}}(0)g^{(2)}_{\hat{o}_2,\,\text{ss}}(0)}.
    \label{CASW0}
\end{equation}
Violation of the first inequality indicates that the emission of photons is anti-correlated, with a diminished chance of a pair being detected together. This occurs in both cavity fields only if their decay rates are sufficiently large. For example, with parameters $g_t = g_c = \gamma$ and $2\kappa_t = 2\kappa_c = \gamma$, one finds $g^{(2)}_{\hat{t},\,\text{ss}}(0)=1.75$ ($g^{(2)}_{\hat{c},\,\text{ss}}(0)=1.86$). However, antibunching is indeed present for larger decay rates; for example, system parameters $g_t = g_c = 4\gamma$ and $2\kappa_t = 2\kappa_c = 16\gamma$ give $g^{(2)}_{\hat{t},\,\text{ss}}(0)=0.083$ ($g^{(2)}_{\hat{c},\,\text{ss}}(0)=0.091$). Despite only being antibunched for the larger decay rate, both arrangements are in clear violation of the second inequality, pertaining to cross-correlations, with $g^{(2)}_{\hat{t}\hat{c},\,\text{ss}}(0)=3.31$ in the first configuration, and $g^{(2)}_{\hat{t}\hat{c},\,\text{ss}}(0)=1.84$ in the second. In both instances, the system therefore acts as a source of quantum light; two-mode nonclassical if the cavity field decay rates are small, and both one- and two-mode nonclassical for larger decay rates.

\begin{figure}[tb]
 	\centering
 	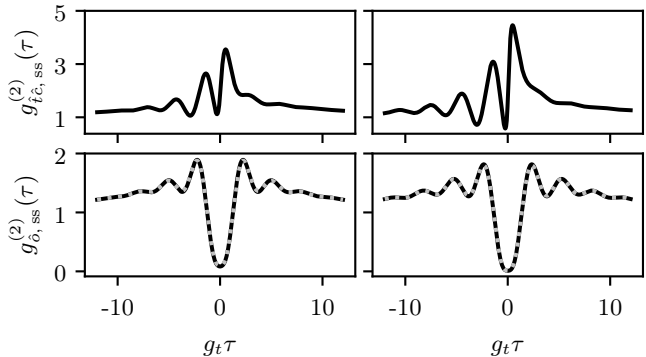
 	\caption{Second-order correlation functions for the steady-state cavity emission, with $g_t = g_c = 4\gamma$, for $\kappa_t=\kappa_c = 8\gamma$ (left column), and $\kappa_t=\kappa_c = 16\gamma$ (right column). The atom has a ground-state splitting $\omega_g = 1000\gamma$, and is symmetrically driven ($\Omega_\text{p} = \Omega_\text{S}$) with an effective two-photon transition rate $\Omega_\text{eff} \approx \Omega_\text{p}\Omega_\text{S}/\Delta_\text{S}=4\gamma$, and $\Delta_\text{p}=-\Delta_\text{S}-\omega_g$. (Top) Cross-correlation, showing conditional likelihood of detecting a photon from the control cavity, given a detection at $\tau=0$ in the telecom cavity. (Bottom) Auto-correlations, which in this case are essentially the same for both cavity modes, with solid black (dotted grey) line showing conditional likelihood of a photon detection from the telecom (control) cavity, given an initial detection from the same mode at $\tau=0$.}
 	\label{fig5}
 \end{figure}

Another element of photon correlations is the time-dependence of the second-order correlation function. This gives a sense of the time-scale over which there is a meaningful change in detection probability, conditioned on a trigger photon detection. One- and two mode Cauchy-Schwarz inequalities also impose classical limitations on the correlation function, relative to the coincidence values. In particular,
\begin{equation}
    g^{(2)}_{\hat{o},\,\text{ss}}(0)\geq g^{(2)}_{\hat{o},\,\text{ss}}(\tau),~
    g^{(2)}_{\hat{o}_1\hat{o}_2,\,\text{ss}}(\tau)\leq \sqrt{g^{(2)}_{\hat{o}_1,\,\text{ss}}(0)g^{(2)}_{\hat{o}_2,\,\text{ss}}(0)}.
    \label{CASWt}
\end{equation}
Fig.~\ref{fig5} shows time-dependent second-order correlations between, and within, the two cavity modes. The structure of the auto-correlations reflects the antibunching that is expected from a cavity coupled to a discrete atomic transition, in the bad-cavity limit ($\kappa\gg g$). Furthermore, the sequential nature of the two transitions manifests itself in the structure of the cross-correlations. A control photon is less likely to be detected in an interval preceding a telecom photon detection than in an interval following it. If the decay rate is sufficiently large, the likelihood of detecting a control photon immediately before a telecom photon is suppressed below the unconditional detection probability, as seen for the larger decay rate in Fig.~\ref{fig5}. Oscillations in the correlation functions are a consequence of the atom coherently evolving through multiple cycles of transitions, each time potentially emitting a telecom photon and then a control photon. The field correlations show violation of the aforementioned time-dependent classical inequalities, further demonstrating that the system is a source of quantum light in steady state.

\section{Alkali Atom Model}
\label{Coupling to the 7s State in Practice}
The conceptual model demonstrates the possibility of extracting quantum-correlated telecom light, in steady state, from a double-diamond model atom. However, implementing such a system with a realistic atom and cavities will naturally introduce additional complications that we consider in this section. We introduce and simulate a model of the hyperfine level structure of a cesium atom, to determine the robustness of the configuration in the face of additional real-world constraints. 

In general, birefringence of the material in a dielectric-guided cavity causes a relative frequency shift between orthogonally-polarized modes of the same index. This splitting is typically sufficiently large that an atom will only interact with a mode of a single linear polarization. This is incorporated into the model with polarization-dependent dipole transition operators, in terms of which the Hamiltonian is expressed in an interaction picture (more details in appendix)
\begin{equation}
	\begin{split}
	\hat{H}/\hbar =&\sum_{nL_J}\sum_{F,m_F}\Delta_{nL_JF}\ket{nL_J,F,m_F}\bra{nL_J,F,m_F}\\
	&+ \frac{\Omega_{\text{p}}}{2}\big(\hat{D}_{\text{p}_\ell}+\hat{D}^\dagger_{\text{p}_\ell}\big)+ \frac{\Omega_{\text{S}}}{2}\big(\hat{D}_{\text{S}_\ell}+\hat{D}^\dagger_{\text{S}_\ell}\big)\\
	 &+ g_t\big(\hat{D}_{{t_\ell}}\hat{t}^\dagger+ \hat{t}\hat{D}^\dagger_{{t_\ell}}\big)+ g_{c}\big(\hat{D}_{{c_\ell}}\hat{c}^\dagger+ \hat{c}\hat{D}^\dagger_{{c_\ell}}\big),\\
	\end{split}
	\label{CsHamiltonian}
\end{equation}
where the sum over $nL_J$ runs over the four relevant fine structure manifolds, and the sum over $F,m_F$ runs over all the hyperfine levels and sublevels for each. The subscript $\ell$ indicates that the laser and cavity fields are linearly polarized along the same direction, parallel to the quantization axis. Co-polarized fields give the highest cavity fluxes as the atomic population tends to accumulate in the hyperfine sublevels with the largest Clebsch-Gordan coefficients for those transitions. The time evolution of the density matrix is governed by a Lindblad master equation:
\begin{equation}
\begin{split}
    \dot{\hat{\rho}} =\frac{1}{i\hbar}\left[\hat{H},\hat{\rho}\right]&+2\kappa_t\mathcal{D}\big(\hat{t}\hspace{0.1em}\big)+2\kappa_c\mathcal{D}\big(\hat{c}\big) \\
    +\sum_q&~ \left[\gamma_\text{p}\mathcal{D}\big(\hat{D}_{\text{p}_q}\big)+\gamma_\text{S}\mathcal{D}\big(\hat{D}_{\text{s}_q}\big) \right. \\[-5pt]
    &~+ \left. \gamma_t\mathcal{D}\big(\hat{D}_{t_q}\big)+\gamma_c\mathcal{D}\big(\hat{D}_{c_q}\big) \right] ,
    \end{split}
\end{equation}
where the sum over $q$ indicates the three independent polarizations of spontaneous emission for each transition, with $q=0$ for $\pi$-polarization and $q=\mp1$ for $\sigma_\pm$-polarization. For a cesium atom, the linewidths of the D-line excited states are $\gamma_\text{p} = 2\pi\cdot 4.575~\text{MHz}$ and $\gamma_\text{c} = 2\pi\cdot 5.234~\text{MHz}$ \cite{SteckCs133}, while the $7S_{1/2}$ state has a total linewidth of $2\pi\cdot 3.296~\text{MHz}$, with  $\gamma_\text{S} = 2\pi\cdot 1.304~\text{MHz}$ and $\gamma_\text{t} = 2\pi\cdot 1.992~\text{MHz}$ from the respective branching ratios \cite{Toh2018,Toh2019}.

\subsection{Choosing the Configuration}
\begin{figure}[htp]
 	\centering
 	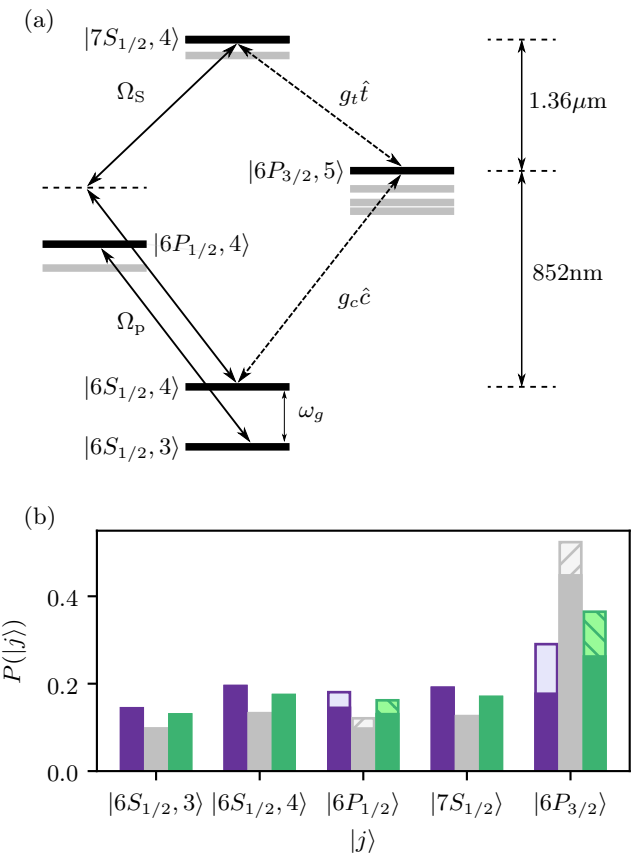
 	\caption{(a) Schematic (not to scale) of the hyperfine levels that are included in the cesium model, with the targeted levels labeled and highlighted in black. (b) Steady-state probability for the atom to occupy state $\ket{j}$, showing the cases of no cavities (purple), just a telecom cavity (grey), and both a telecom and a control cavity (green). In all cases, the driving is symmetric ($\Omega_\text{p} = \Omega_\text{S} \approx 2\pi\cdot 1450~\text{MHz}$) and, where included, the cavity modes have a maximum coupling strength $g_j = 1.5\times4(\gamma_\text{S} +\gamma_t) \approx 2\pi\cdot 19.8~\text{MHz}$, and decay rate $\kappa_t = \kappa_c = 8(\gamma_\text{S} +\gamma_t)\approx 2\pi\cdot 26.4~\text{MHz}$. Note that the cavity couplings are scaled by a factor of $1.5$ compared to the simple model, to normalize for the maximal Clebsch-Gordan coefficient for $\pi$-polarized transitions. Dark colors show the fraction of population within each state that occupies the targeted hyperfine levels, and lighter colors indicate the stray population in other hyperfine levels. In all cases, the total fraction of population within the targeted levels is $>0.85$.}
 	\label{CsFineLevels}
 \end{figure}

In order to reach a non-trivial steady state, it is necessary to carefully choose parameters and energy levels that ensure the desired sets of transitions occur sufficiently quickly, whilst avoiding non-targeted hyperfine levels. We consider a single cesium atom, and the results are similar regardless of which ground state is chosen, so results are only presented for the case where the pump laser is near-resonant from the $F=3$ ground state, and the two-photon transition is resonant from the $F=4$ ground state, as depicted in Fig.~\ref{CsFineLevels}(a). The specific energy levels that are targeted in this section are listed as follows, with the corresponding representative level in the conceptual model,
\begin{equation}
	\begin{split}
    \ket{g_1}&\mapsto\ket{6S_{1/2},3}
    ,~\ket{g_2}\mapsto\ket{6S_{1/2},4},\\
    \ket{e_1}&\mapsto\ket{6P_{1/2},4},~
    \ket{e_2}\mapsto\ket{6P_{3/2},5},\\
    \ket{f}&\mapsto\ket{7S_{1/2},4},\\
   \end{split}
\end{equation}
where we use the shorthand notation
\begin{equation}
\begin{split}
\ket{nL_{J},F}\equiv\sum_{m_F}\ket{nL_{J},F,m_F}.
\end{split}
\end{equation}
These targeted levels are shown in black and explicitly labeled in Fig. \ref{CsFineLevels} (a), for comparison with the respective levels in the conceptual model. 

A caveat to the choice of using all $\pi$-polarized coupling fields is that the $\ket{6S_{1/2},3,0}\leftrightarrow\ket{6P_{1/2},3,0}$ transition is electric-dipole forbidden, so the pump laser is arranged to resonantly drive $\ket{6S_{1/2},3}\leftrightarrow\ket{6P_{1/2},4}$ transitions. It is a natural choice to use the $\ket{6P_{3/2},5}$ level as the second intermediate state, given that it can only decay to the uppermost ground state, as is desired. To facilitate this choice, the Stokes laser is tuned so that the two-photon transition is resonant to the $\ket{7S_{1/2},4}$ level. The ground state splitting in cesium is almost an order of magnitude larger than the $6P_{1/2}$ hyperfine splitting, so the two-photon transition proceeds collectively through both of the hyperfine levels. While this is critical for the scheme to operate efficiently, given that the $\ket{nS_{1/2},4,0}\leftrightarrow\ket{6P_{1/2},4,0}$ transitions are electric-dipole forbidden, the excitation pathways in fact interfere destructively. Strong laser fields are therefore needed to ensure an appreciable two-photon excitation; that is, we require
\begin{equation}
    \Omega_p\Omega_s \left(\frac{1}{\omega_g}-\frac{1}{\omega_g+\omega_{D_1}}\right)\gg \gamma_S + \gamma_t,
\end{equation}
where $\omega_{D1}$ is the $6P_{1/2}$ hyperfine splitting.

Choosing the target energy levels as discussed, Fig.~\ref{CsFineLevels}(b) compares the steady-state atomic populations for an atom in free space, and when coupled to one or two cavity modes. As expected from the idealized model, coupling a telecom cavity mode to the atom yields a larger population in the $6P_{3/2}$ states, as the cycle of transitions is limited by spontaneous emission on the D2 line. Adding the control cavity expedites this transition, and allows for a further increase in the telecom photon flux. For the driving parameters in Fig.~\ref{CsFineLevels}(b), and without coupling to any cavity modes, the atom emits telecom light at a rate of $0.19\gamma_t$ in steady state. Under the same conditions, coupling the atom to a telecom cavity mode yields an output flux of $\Phi^t_{\text{ss}} = 0.44\gamma_t$, while the addition of a coupling cavity increases this to $\Phi^t_{\text{ss}} = 0.60\gamma_t$, for the cavity parameters used. This corresponds to a telecom cavity photon emission rate of $1.20~\text{MHz}$ ($0.87~\text{MHz}$) for the system with (without) the control cavity.

\subsection{Second-Order Correlations}
\label{Nonclassical Photon Statistics at a Telecom Wavelength}
For the parameters chosen in Fig.~\ref{CsFineLevels} (b), the telecom and control cavity output fields show strongly anti-correlated photon counting statistics. With just the telecom cavity coupled to the atom, the telecom mode has a second-order coincidence likelihood of $g^{(2)}_{\hat{t},\text{ss}}(0) = 0.057$. When coupled to both cavities, one finds $g^{(2)}_{\hat{t},\text{ss}}(0) = 0.097$ and $g^{(2)}_{\hat{c},\text{ss}}(0) = 0.110$. Furthermore, the cross-correlation between cavity modes takes an initial value of $g^{(2)}_{\hat{t}\hat{c},\text{ss}}(0) = 1.714$. In all cases, these auto- and cross-correlations are clearly in violation of the Cauchy-Schwarz inequalities (\ref{CASW0}).

\begin{figure}[htb]
 	\centering
 	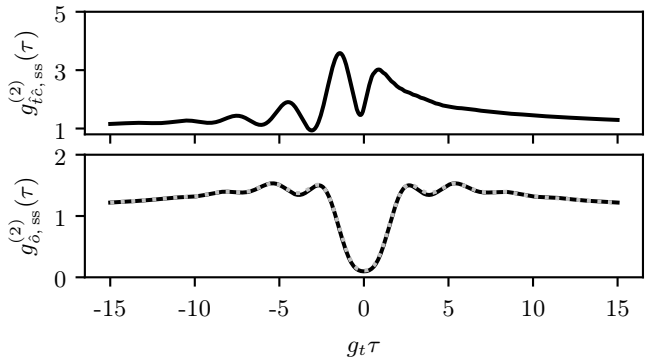
 	\caption{Second-order correlation functions for the steady-state cavity emission, under symmetric driving ($\Omega_\text{p} = \Omega_\text{S} \approx 2\pi\cdot 1450~\text{MHz}$), and with maximal coupling strength $g_j = 1.5\times 4(\gamma_\text{S} +\gamma_t) \approx 2\pi\cdot 19.8~\text{MHz}$, and decay rate $\kappa_t = \kappa_c = 8(\gamma_\text{S} +\gamma_t)\approx 2\pi\cdot 26.4~\text{MHz}$. (Top) Cross-correlation, showing conditional likelihood of detecting a photon from the control cavity, given a detection at $\tau=0$ in the telecom cavity. (Bottom) Auto-correlations, which are essentially the same for both cavity modes, with solid black (dotted grey) line showing conditional likelihood of a photon detection from the telecom (control) cavity, given an initial detection from the same mode at $\tau=0$.}
 	\label{CsCorrelations}
 \end{figure}

Time-dependent, second-order auto- and cross-correlation functions are shown in Fig.~\ref{CsCorrelations}. The structure of these correlations is similar to those in the conceptual model. In particular, the cross-correlation confirms that one is more likely to detect a control photon immediately after a telecom photon, compared to in the reverse ordering. The oscillations in the correlation function are not as pronounced as for the ideal model, due to the additional spontaneous emission pathways. However, the fact that they persist shows the system coherently undergoing the desired cycle of transitions. Both fields show a tendency for temporal delay between successive photon emissions within the same mode, as is characteristic of an atomic transition. The correlations indeed violate the Cauchy-Schwarz inequalities \ref{CASWt}, corroborating the fact that the system behaves as a steady-state source of quantum-correlated light in the telecom band. 

\section{Conclusion}
We have investigated the steady-state generation of quantum-correlated, telecom-band light from a single atom. By appropriately configuring a pair of lasers to excite the atom via both a one- and two-photon resonance, a cycle of transitions is established that continually transfers light into a telecom-band cavity mode. Antibunched photon counting statistics are demonstrated in the output field, which reflects the discrete nature of the underlying atomic transition. Additionally, we show that adding a control cavity can enhance the steady-state rate of telecom photon emission. The second cavity also enables the generation of nonclassical two-mode correlations between the two cavity output fields. While preparing this manuscript, we became aware of related work that addresses the pulsed generation of photon pairs from a doubly-resonant Fabry-P\'erot cavity, using a telecom-band transition in ${}^{87}$Rb \cite{Ruskuc2025}. The cavity architecture developed therein presents a another promising avenue to pursue the steady-state schemes developed in this work.

\section*{Acknowledgments} 
AE thanks Quantum Technologies Aotearoa (QTA) for supporting his visit to the group of TA at Waseda University as part of this research. SP also thanks the group of TA for support and hospitality during a visit to Waseda University.

\bibliography{References.bib}
\appendix

\onecolumngrid

\section{Interaction Picture Hamiltonian For the Cesium Atom}
\label{appendix}
\noindent
We first outline the Hamiltonian in the Heisenberg picture, starting with the bare cavity and atomic energies
\begin{equation}
	\hat{H}_0/\hbar = \omega_t\hat{t}^\dagger\hat{t} + \omega_c\hat{c}^\dagger\hat{c}+ \sum_{nL_J}\sum_{F,m_F}\omega_{nL_JF}\ket{nL_J,F,m_F}\bra{nL_J,F,m_F},
\end{equation}
where the sum over $nL_J$ spans the four fine-structure manifolds included in the model; $6S_{1/2}$, $6P_{1/2}$, $6P_{3/2}$, and $7S_{1/2}$. The hyperfine energy levels, $\omega_{nL_JF}$ are all specified relative to the zero-point energy at the $\ket{6S_{1/2},F=4}$ state, while $\omega_t$ and $\omega_c$ are the bare resonances of the telecom and control cavities, respectively.

Transitions from one fine-structure level to another are treated collectively with respect to frequency, but independently with respect to polarization. We define electric-dipole transition operators with Clebsch-Gordan coefficients written in terms of the Wigner-3j and -6j symbols, as defined in \cite{Brink1994}:

\begin{equation}
		\begin{split}
		\hat{D}_q(nL_J,n'L'_{J'}) =\sum_{F,F'}\sum_{m_F,m_{F'}}& (-1)^{J+I-m_F}\sqrt{(2F+1)(2F'+1)(2J'+1)}\\&\times\begin{pmatrix}
			F' & 1 & F \\
			m_F' & q & -m_F 
		\end{pmatrix}\begin{Bmatrix}
		J & J' & 1 \\
		F' & F & I 
		\end{Bmatrix}\ket{nL_J,F,m_F}\bra{n'L'_{J'},F',m_{F'}},
		\end{split}
		\label{DipoletransitionOperator}
	\end{equation}
where $q=0$ for $\pi$-polarization, and $q=\mp1$ for $\sigma_\pm$-polarization. For brevity in text, we label operators of a particular polarization for each transtions:
\begin{equation}
\begin{split}
    \hat{D}_{\text{p}_q}\equiv\hat{D}_{q}(6S_{1/2},6P_{1/2}),~\hat{D}_{\text{S}_q}\equiv\hat{D}_{q}(6P_{1/2},7S_{1/2}),\\
    \hat{D}_{t_q}\equiv\hat{D}_{q}(6P_{3/2},7S_{1/2}),~\hat{D}_{c_q}\equiv\hat{D}_{q}(6S_{1/2},6P_{3/2}).
    \end{split}
\end{equation}
In terms of these operators, the dipole coupling Hamiltonian for the two driving lasers is expressed concisely as
\begin{equation}
	\hat{H}_\text{lasers} =  \frac{\Omega_{\text{p}}}{2}\left(\hat{D}_{\text{p}_\ell}e^{i\omega_\text{p}t}+ \hat{D}_{\text{p}_\ell}^\dagger e^{-i\omega_\text{p}t}\right)+  \frac{\Omega_{\text{S}}}{2}\left(\hat{D}_{\text{S}_\ell}e^{i\omega_\text{S}t}+ \hat{D}_{\text{S}_\ell}^\dagger e^{-i\omega_\text{S}t}\right),
\end{equation}
where $\Omega_{\text{p}}$ ($\Omega_{\text{S}}$) and $\omega_\text{p}$ ($\omega_\text{S}$) are the Rabi frequency and bare frequency of the ${q}$-polarized pump (Stokes) laser. As mentioned in the text, the subscript $\ell$ indicates a co-linear polarization that is aligned parallel to the quantization axis, corresponding to $\pi$-polarization (i.e. $q=0$). The two cavity couplings are described by the Hamiltonian 
\begin{equation}
	\hat{H}_\text{cavities} = g_t\left(\hat{D}_{t_\ell}\hat{t}^\dagger+ \hat{t}\hat{D}_{t_\ell}^\dagger\right)+g_c\left(\hat{D}_{c_\ell}\hat{c}^\dagger+\hat{c} \hat{D}_{c_\ell}^\dagger\right).
\end{equation}
Defining the transformation Hamiltonian,
\begin{equation}
	\begin{split}
	\hat{H}_0' =& (\omega_\text{p}+\omega_\text{S} - \omega_{c})\hat{t}^\dagger\hat{t} + \omega_{c}\hat{c}^\dagger\hat{c}+\omega_\text{p}\sum_{F,m_F}\ket{6P_{1/2},F,m_F}\bra{6P_{1/2},F,m_F} \\&+ \omega_{c}\sum_{F,m_F}\ket{6P_{3/2},F,m_F}\bra{6P_{3/2},F,m_F} + (\omega_\text{p}+\omega_\text{S})\sum_{F,m_F}\ket{7S_{1/2},F,m_F}\bra{7S_{1/2},F,m_F} 
	,
	\end{split}
\end{equation}
and performing a unitary transformation into the corresponding interaction picture, one arrives at the Hamiltonian \ref{CsHamiltonian} by making the rotating-wave approximation, with the resonance assumption that $\omega_\text{p} + \omega_\text{s} = \omega_t + \omega_c$. The detunings of the target atomic states in the interaction picture are specified as $\Delta_{6P_{1/2}F} =\omega_{6P_{1/2}F} - \omega_\text{p} $, $\Delta_{6P_{3/2}F} = \omega_{6P_{3/2}F}-\omega_{c}$, and $\Delta_{7S_{1/2}F} = \omega_{7S_{1/2}F} - (\omega_\text{p} + \omega_\text{S})$. Choosing the laser frequencies appropriately, the one-photon resonance condition gives $\Delta_{6P_{1/2}4}=-\omega_g$, while the two-photon resonance condition gives $\Delta_{7S_{1/2}4}=0$. The control cavity resonance is arranged such that $\Delta_{6P_{3/2}5}=0$.
\end{document}